\documentclass[aps%
 ,secnumarabic%
,preprint
,amssymb, amsmath,nobibnotes, aps, pre, floatfix,superscriptaddress]{revtex4-1}
\usepackage{graphicx,epsfig}
\usepackage{amsmath}
\usepackage{amsfonts}
\usepackage{fancyhdr}
\usepackage{hyperref}
\usepackage{dirtytalk}
\usepackage{epstopdf}
\usepackage{longtable}
\usepackage{cleveref}
\usepackage{gensymb}
\begin{document}
\pagestyle{fancy}
\lhead{}
\rhead{V. Navas-Portella, I. Serra, \'A. Corral, and E. 
Vives}
\lfoot{}
\rfoot{}
%%%%%%%%%%%%%%%%%%%%%%%%%%%%%%%%%%%%%%
\title{Increasing power-law %% millor stressar power law
range in avalanche amplitude and energy distributions}
\author{V\'ictor Navas-Portella}
\affiliation{Centre de Recerca Matem\`atica, Edifici C, Campus Bellaterra, E-08193 Bellaterra, Catalonia, Spain.}
\affiliation{Barcelona Graduate School of Mathematics, Edifici C, Campus Bellaterra,
E-08193 Barcelona, Spain} 
\affiliation{Facultat de Matem\`atiques i Inform\`atica, Universitat de Barcelona, Barcelona, Spain} 
\author{Isabel Serra}
\affiliation{Centre de Recerca Matem\`atica, Edifici C, Campus Bellaterra, E-08193 Bellaterra, Catalonia, Spain.}
\author{\'Alvaro Corral}
\affiliation{Centre de Recerca Matem\`atica, Edifici C, Campus Bellaterra, E-08193 Bellaterra, Catalonia, Spain.}
\affiliation{Barcelona Graduate School of Mathematics, Edifici C, Campus Bellaterra,
E-08193 Barcelona, Spain} 
\affiliation{Complexity Science Hub Vienna,
Josefst\"adter Stra$\beta$e 39,
1080 Vienna,
Austria}
\affiliation{Departament de Matem\`atiques, Universitat Aut\`onoma de Barcelona,
E-08193 Barcelona, Spain}
\author{Eduard Vives}
\affiliation{Departament de Mat\`eria Condensada, Facultat de F\'{\i}sica, Universitat de Barcelona, Mart\'i Franqu\`es 1, 08028 Barcelona, Catalonia, Spain.}
\affiliation{Universitat de Barcelona Institute of Complex Systems (UBICS), Facultat de F\'{\i}sica, Universitat de Barcelona, Barcelona, Catalonia, Spain.}
\date{\today}
%\email{vicnavas1992@gmail.com} %optional
%\date{\today}
\begin{abstract}
Power-law type probability density functions spanning several orders of magnitude are found for different avalanche properties. We propose a methodology to overcome empirical constrains that limit the power-law range for the distributions of different avalanche observables like amplitude, energy, duration or size. By considering catalogs of events that cover different observation windows, maximum likelihood estimation of a global power-law exponent is computed. This methodology is applied to amplitude and energy distributions of acoustic emission avalanches in failure-under-compression experiments of a nanoporous silica glass, finding in some cases global exponents in an unprecedented broad range: $4.5$ decades for amplitudes and $9.5$ decades for energies. In the later case, however, strict statistical analysis suggests experimental limitations might alter the power-law behavior.
\end{abstract}
\maketitle
%%%%%%%%%%%%%%%%%%%%%%%%%%%%%%%%%%%%%%%%%%%%%%%%%%%%%%%%%%%%%%%%%%%%%%%%%%%
\section{Introduction}
Avalanche processes are present in a vast number of out-of-equilibrium physical phenomena \citep{Sethna2001,Salje2014a,Bak1996}. These processes are characterized by intense bursts of activity preceded by periods of silence. Some properties that characterize this kind of processes can be described in terms of probability density functions (PDFs) that exhibit lack of finite moments due to their power-law shape. Consequently, fitting PDFs to different avalanche properties such like sizes, energies, or amplitudes is a task that requires a rigorous treatment \citep{Clauset2009,Deluca2013}. 
One of the most important features of this kind of functions is their invariance under any scale transformation. This property of scale invariance can be written as $f \left( \lambda x \right) = \lambda^{-\gamma}f\left( x \right)$ for $x \in \left(0,+\infty \right)$. The only solution for all $\lambda$ of this functional equation \citep{Christensen_Moloney} is a power-law $f(x)=k x^{-\gamma}$, where the exponent $\gamma$ can take any real value and $k$ is a constant.

Some experimental works confirm the presence of scale invariance in data by assuming power-law 
behavior for which
data scarcely covers few orders of magnitude \citep{Friedman2012,Papanikolaou2011,Uhl2015}. The broader the distribution range the more reliable the property of scale invariance in experimental data.
Amplitude distributions have been studied in different experimental works based on the amplitude of acoustic emission (AE) avalanches \citep{Vives1994,Petri1994,Vives1995,Carrillo1998,Weiss2007}. However, experimental fitted distributions expand at most two orders of magnitude in voltages \citep{Vives1994,Weiss2000,Weiss2001,Weiss2007} due to the limitations in the observation windows. Typically the existence of noise and/or under-counting effects affects the smallest observable values, whereas saturation and/or lack of statistics due to under-sampling limits the largest observable values. In most cases, these experimental limitations are not sharp due to electronic uncertainties. Recent studies regarding the AE in compression experiments of porous materials \citep{Baro2013,Nataf2014,Navas-Portella2016}, wood \citep{Makinen2015}, ethanol-dampened charcoal \citep{Ribeiro2015}, confined-granular
matter under continuous shear \citep{Lherminier2015}, etc. have focused the attention in the energy 
distribution of avalanches due to the similarities with the Gutenberg-Richter law for earthquakes \citep{Serra_Corral}. 

In this work, we provide a procedure to broaden the range of validity of power-law like behavior of the distributions corresponding to avalanche amplitudes and energies. From a set of $n_{\rm cat}$ catalogs of events whose measured properties span different observation windows, data analysis is performed by assembling them in order to obtain global exponents that characterize the distribution of these avalanche properties. Through this procedure, the fitted global distribution spans a broader range than the one from the fit of every individual catalog.

This manuscript is organized as follows: In Section \ref{sec:methodology} an overview of the fitting procedure is shown. In Section \ref{sec:vycor} we present the experimental methodology in the recording of AE during displacement-driven compression of porous glasses \citep{Navas-Portella2016}. In Sections \ref{seq:amplitudes}   and \ref{seq:energy} avalanche amplitudes and energies are studied respectively by applying the methodology exposed in Sec. \ref{sec:methodology}. Finally, a brief summary of the results will be presented in Sec. \ref{Conclusions}.
\section{General Methodology}
\label{sec:methodology}
By considering $n_{\rm cat}$ catalogs of $N_{i}$ ($i=1,...,n_{\rm cat}$) events each, corresponding to different experiments (or different observation windows) and characterized by a set of variables (amplitude, energy, duration, etc.), one wants to fit a general power-law type PDF with a global exponent for all the catalogs. Note that, in  the $i$-th catalog, the variable $\mathcal{X}$ can acquire values in a range typically spanning several orders of magnitude. 
The first step consist in fitting a power-law PDF in a range $\left[ a_{i},b_{i} \right]$ for each catalog via maximum likelihood estimation (MLE) and goodness-of-fit testing \citep{Deluca2013}. Details of this fitting-procedure are explained in Appendix \ref{sec:AP1}. By this method we correct for problems close to the limits of the observation windows although discarding some experimental data. In this situation, one may be able to state that, for the $i$-th catalog, the variable $\mathcal{X}$ follows a power-law PDF $f^{(i)}_{\mathcal{X}}(x; \hat{\gamma}_{i},a_{i},b_{i})$ in a certain range $\left[a_{i},b_{i}\right]$ with exponent $\hat{\gamma}_{i}$ and a number $\hat{n}_{i}$ of data entering into the fit $\left(\hat{n}_{i} \leq N_{i}\right)$. Under these conditions, the next null hypothesis $\rm H_{0}$ is formulated: the variable $\mathcal{X}$ is power-law distributed with a global exponent $\Gamma$ for all the catalogs. 

The log-likelihood function of this global distribution can be written as:
\begin{equation}
\log \mathcal{L}= \sum_{i=1}^{n_{\rm cat}} \sum_{j=1}^{\hat{n}_{i}} \log f^{(i)}_{\mathcal{X}} \left( x_{ij}; \Gamma, a_{i},b_{i} \right)
\label{eq:general}
\end{equation}
 where $x_{ij}$ corresponds to the values of the variable $\mathcal{X}$ in the $i$-th catalog, $\hat{n}_{i}$ is the number of data  between $a_{i}$ and $b_{i}$ in the $i$-th catalog and $\Gamma$ is the global exponent. Since the particular ranges $\left[ a_{i},b_{i} \right]$ and the number of data $\hat{n}_{i}$ are known, one has to find the value of the exponent $\hat{\Gamma}$ that maximizes the log-likelihood expression in Eq. (\ref{eq:general}).

Intuitively, one could be tempted to think that the null hypothesis will not be rejected if the values of the particular exponents $\gamma_{i}$ do not differ too much. Nevertheless, a more rigorous treatment is required. Statistical procedures, such as a permutational test \citep{Deluca2016}, could be used in order to check whether the exponents are the same or not. However, since we propose a global distribution characterized by a global exponent $\hat{\Gamma}$, a goodness-of-fit test for this global distribution is performed in order to determine whether the null hypothesis can be rejected or not \citep{Deluca2013,Clauset2009}. If the goodness-of-fit test yields a high enough $p$-value, one is able to state that the variable $\mathcal{X}$ is power-law distributed with exponent $\hat{\Gamma}$ along all the different catalogs or experiments, with ranges $\left[ a_{i},b_{i} \right]$ each.  Details of the goodness-of-fit test are exposed in Appendix \ref{sec:AP4}. In this way, if these intervals span different orders of magnitude,  one can increase the power-law range in several decades. An alternative procedure, where the ranges $\left[ a_{i},b_{i} \right]$ are optimized directly from Eq. (\ref{eq:general}), is disregarded for being enormously computer-time consuming.
 
This methodology is applied to avalanche amplitudes and energies on AE data in failure-under-compression experiments of nanoporous silica glasses. Since the experimental set-up records discrete values for the amplitude (in  dB) and almost continuous values of the energy (in aJ), particular expressions for the log-likelihood Eq. (\ref{eq:general}) as well as the different ways of implementing the goodness-of-fit test will be explained in Sections \ref{seq:amplitudes} and \ref{seq:energy}.
\section{Failure under compression of porous glasses}
\label{sec:vycor}
Uni-axial compression experiments of porous glass Vycor (a nanoporous silica glass with $40\%$ porosity) are performed in a conventional test machine ZMART.PRO (Zwick/Roell). Cylindrical samples with no lateral confinement are placed between two plates that approach each other at a certain constant rate $\dot{z}$. We refer to such a framework as displacement-driven-compression. With the aim of having the same conditions for all the experiments, samples have the same diameters $\Phi = 4.45$mm and heights $H=8$mm, and the compression rate is fixed at $\dot{z}=0.005$mm/min. Before compression, samples were cleaned with a $30\%$ solution of $\rm H_{2}O_{2}$, during 24 h and dried at 130$^{\circ}$C.  Simultaneous to the compression, recording of an AE signal is performed by using a piezoelectric transducer embedded in one of the compression plates. The electric signal $U(t)$ is pre-amplified, band filtered (between 20 kHz and 2 MHz), and analysed by means of a PCI-2 acquisition system from Euro Physical Acoustics (Mistras Group) with an AD card working at 40 Megasamples per second with 18 bits precision \citep{PCI2}. This should be kept in mind when considering some of the measures as continuous (energy or voltage). Recording of data stops when a big failure event occurs and the sample gets destroyed.

We prescribe that an AE avalanche event (often called AE hit in specialized AE literature) starts at the time $t_{j}$ when the signal $U(t)$ crosses a fixed detection threshold and finishes at time $t_{j}+\tau_{j}$ when the signal remains below threshold from $t_{j}+\tau_{j}$ to at least $t_{j}+\tau_{j}+200\mu$s. The amplitude $A$ recorded in  dB follows the expression $A=\left[20 \log_{10}\left( \vert V \vert / V_{0} \right) \right]$, where $V$ is the peak voltage achieved by the AE signal during the event, $V_{0}= 1  \mu$V is a reference voltage, and the brackets round the value to its nearest integer in  dB. Such a procedure is extensively used in electronic systems.
Note that in our terminology $A$ will be called amplitude in  dB,
whereas the peak voltage $V$ will be refered to simply as amplitude,
in agreement with previous literature.
From the values of $A$ one can obtain the values $y$ of the discretized peak-voltage:
\begin{equation}
y=g(A)=V_{0}10^{A/20}.
\label{eq:canvia}
\end{equation}
As the values of $A$ are integer, 
the values of $y$ will no longer be integer but they will collapse into a set of values $\lbrace y_{1},y_{2},...,y_{j},...,y_{k}  \rbrace$ measured in $\mu V$.
The energy $E_{j}$ of each avalanche or event is determined as $E_{j}=\frac{1}{R}\int_{t_{j}}^{t_{j}+\tau} U^{2}(t) dt$ where $R$ is a reference resistance of $10$ k$\Omega$. At the end of one experiment, one has a catalog or collection of events each of them characterized by a time of occurrence $t$, amplitude in  dB $A$, energy $E$, and duration $\tau$.

In order to obtain catalogs that span different observation windows, displacement-driven compression experiments with Vycor cylinders have been performed for different values of the pre-amplification and the detection threshold. In this case, $n_{\rm cat}=4$ experiments have been performed with the following pre-amplification values: 60  dB, 40 dB, 20 dB and 0 dB, and the respective values of the detection threshold 23 dB, 43 dB, 63 dB and 83 dB referring to the signal $U(t)$, not the preamplified signal (in such a way that after preamplification the threshold always moves to $83$ dB). This value of the threshold is as low as possible in order to avoid parasitic noise.

Signal pre-amplification is necessary if one wants to record small AE events. Some values of the pre-amplified signals are so large that can not be detected correctly by the acquisition system. This fact leads to a saturation in the amplitude and, consequently, an underestimated energy of the AE event. This effect can be immediately observed in the distributions of the amplitude in  dB, Fig. \ref{fig:fig1}, where there is an excess of AE events in the last bin of amplitudes for the experiments at $60$ dB and $40$ dB. Note that the thresholding we perform turns out to be of the same kind as that in Refs.\citep{Deluca2015,Font-Clos2015}.
\section{Amplitudes}
\label{seq:amplitudes}
\begin{figure}
\includegraphics[scale=0.75]{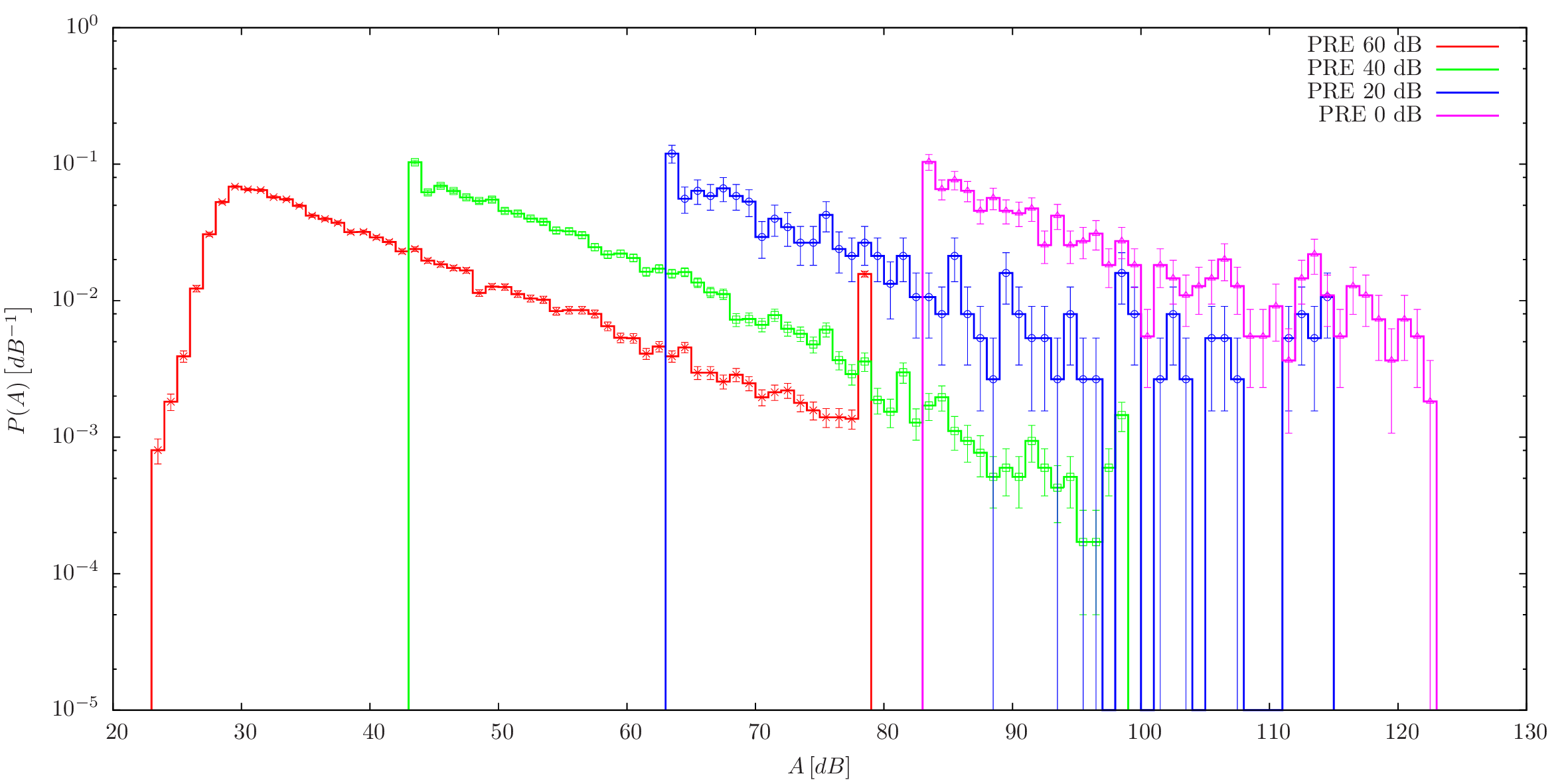}
\caption{\label{fig:fig1} Estimated probability mass functions of the amplitude in  dB for the complete datasets of the different experiments performed at different pre-amplifications (PRE). Error bars are estimated as the standard deviation for each bin \citep{Deluca2013}. }
\end{figure}

In Fig. \ref{fig:fig1} we show the probability mass functions of the amplitude in  dB for the complete datasets of all the experiments performed at different values of the pre-amplification. 
\subsection{Particular fits}

We consider that, for each experiment, the random variable $\mathcal{V}$
corresponding to amplitude 
(i.e., the peak voltage, whose values are denoted by $V$)
follows a truncated continuous power-law distribution,
\begin{equation}
f_{\mathcal{V}}(V) dV=\frac{1-\alpha}{V_{\rm max}^{1-\alpha}-V_{\rm min}^{1-\alpha}}V^{-\alpha} dV.
\label{eq:ampdis}
\end{equation}
However, the true value of $\mathcal{V}$ is not accessible from the experiments, 
and what we have instead is its discretized counterpart $\mathcal{Y}$
(the discretized peak voltage, whose values are denoted by $y$),
which is  concentrated in $k$ discrete values (but not equispaced).
%
%Regarding data, we can consider the recorded amplitude $\mathcal{Y}$ as a random variable concentrated in those $k$ discrete values distributed as a truncated power-law. We will perform the fit by assuming that the continuous random variable $\mathcal{V}$ (whose exact values are not recorded) follows a continuous truncated power-law distribution:
%\begin{equation}
%f_{\mathcal{V}}(V) dV=\frac{1-\alpha}{V_{\rm max}^{1-\alpha}-V_{\rm min}^{1-\alpha}}V^{-\alpha} dV
%\label{eq:ampdis}
%\end{equation}
In fact, the values $V$ that the variable $\mathcal{V}$ 
can take are the real values of the voltages read by the AD card, but they are transformed into  dB, losing precision. 

Under the assumption of a power-law distributed $\mathcal{V}$, 
we are able to state that the variable $\mathcal{Y}$ has probability mass function
$$
f_{\mathcal{Y}}\left( y\right) = P\left( \mathcal{Y}=  y  \right) =
P \left[ g\left(A-\Delta\right) \leq \mathcal{V} < g\left(A+\Delta \right) \right]
$$
\begin{equation}
= \frac{ g^{1-\alpha}\left( A + \Delta \right) - g^{1-\alpha}\left( A - \Delta \right) }{V_{\rm max}^{1-\alpha}-V_{\rm min}^{1-\alpha}}
=\frac{2 \sinh [2.30(1-\alpha) \Delta / 20]}
{V_{\rm max}^{1-\alpha}-V_{\rm min}^{1-\alpha}} \, \frac 1 {y^{\alpha-1
}},
\label{eq:conc}
\end{equation}
where $y=g\left( A \right)$, 
$\Delta =0.5$  dB is a vicinity around the values of $A$, 
$2.30 \simeq \log 10$,
and 
$P$ refers to a probability. 
Note that $f_{\mathcal{Y}}\left( y\right)$ is a power law
but with exponent $\alpha-1$.
%One can obtain the values of $y$ through Eq. (\ref{eq:canvia}) implying that the fit of a truncated power-law is done to data which are concentrated in $k$ values of $\mathcal{Y}$. 

The log-likelihood function for a particular experiment can be written as:
\begin{equation}
\log \mathcal{L} = \sum_{\rm l= A_{min}}^{\rm A_{max}} \omega_{l} \log  f_{\mathcal{Y}}\left( g(l) \right)  = 
  \sum_{\rm l= A_{min}}^{\rm A_{max}} \omega_{l} \log P \left( g\left(l-\Delta\right) \leq \mathcal{V} < g\left(l+\Delta \right)  \right) 
\label{eq:mleamp}
\end{equation}
where $\rm A_{min}$ and $\rm A_{max}$ are the values of the amplitude in  dB corresponding to the cutoffs imposed on the sample for the analysis (see Appendix \ref{sec:AP1} for further details). The frequency $\omega_{l}$ is the number of events with discretized peak voltage $y_{l}=g(l)$. 
The next step consists in finding the value of $\alpha$ that maximizes Eq. (\ref{eq:mleamp}) using a numerical method. The values of the fitted exponent $\alpha$ for different values of $A_{\rm min}$ and $A_{\rm max}$ are shown in Appendix \ref{sec:AP3} by using MLE exponent maps \citep{Baro2012}. Once the exponent is found, one has to determine whether the fit is appropriate to data or not. All the details concerning the fitting procedure and the statistical test are exposed in Appendix \ref{sec:AP1}.

In Table \ref{tab:1} we present the fitting values of the particular fits: exponents $\hat{\alpha}$, ranges $\left[ A_{\rm min}, A_{\rm max}  \right]$, number of events $\hat{n}$ included in the fit and an estimated $p$-value. Numbers in parenthesis in the columns specifying the ranges $\left[ A_{\rm min}, A_{\rm max} \right]$ correspond to the total range of the sample. Each experiment detects avalanches within $2.8$ decades in amplitude but all the experiments together would yield a total range of $5.8$ decades. This range is broader than other ranges of AE amplitudes \citep{Petri1994,Vives1995,Carrillo1998,Weiss2000,Koslowski2004,Richeton2005,Weiss2007}.
It must be mentioned that performing these particular fits by simply assuming that the discrete variable $\mathcal{Y}$ directly follows a truncated continuous power-law leads to the rejection of this hypothesis in the goodness-of-fit test. 

According to the range of detection for the experiment performed at 0  dB, one could be able to observe events up to $139$  dB. Nevertheless, the maximum in this sample corresponds to $123$  dB. Under the hypothesis that a power-law distribution with the same exponent  ($\hat{\alpha}=1.61 \pm 0.04$) can be extended for larger values of the amplitude, the probability $P \left( 123 \rm{ dB} < A < 140 \rm{ dB} \right)$ turns out to be  $P=0.042$. For $N_i=548$ trials, the probability of having no events in this range can be estimated by $\left( 1-0.042 \right)^{548}= 6.2 \times 10^{-11}$. Based on these simple calculations, one could justify the existence of a corner value due to the finite size of the sample properties \citep{Serra_Corral}. 
However, this corner value would be only visible for the experiment at zero amplification;
the same calculation for the experiment at 20 dB gives a probability of having no events
above the maximum observed of 0.14, 
which is not an extremal value at all.
\begin{table}[htbp]
\begin{tabular}{| l | r | rr | rr | rr | r |c| }
\hline
\multicolumn{1}{| l |}{PRE in dB} & \multicolumn{1}{| c |}{$\hat{\alpha}$} & \multicolumn{2}{| c |}{$A_{\rm min}$ in dB}  & \multicolumn{2}{| c |}{$A_{\rm max}$ in dB} & \multicolumn{2}{| c |}{$\hat{n}(N)$} & \multicolumn{1}{| c |}{$p$-value}  \\ \hline
60 & $1.743\pm 0.007$  & $32$ & $(23)$ & $78$ & $(79)$ & 21414 &(28614) & $0.92$  \\ \hline
40 & $1.75 \pm 0.01$ & $46$ & $(43)$ & $72$ & $(99)$ & 9146 &(11717) & $0.20$  \\ \hline
20 & $1.67\pm 0.04$ & $64$ & $(63)$ & $114$ & $(115)$ & 353 &(376) & $0.50$ \\ \hline
0 & $1.61\pm 0.04$ & $84$ & $(83)$ & $122$ & $(123)$ & 528 &(548) & $0.64$ \\ \hline
\textbf{Global} & 1.740 $\pm$ 0.006 & $32$ & $(23)$ & $ 122$ & $(123)$ & 31441 & (41255) & 0.36 \\ \hline
\end{tabular}
\caption{Fitted parameters for Eq. (\ref{eq:ampdis}) for each particular experiment and for the global fit. $\hat{\alpha}$ corresponds to the fitted exponent in the range $\left[ A_{\rm min},A_{\rm max} \right]$ for which the goodness-of-fit test exceeds the significance level $p_{c}=0.2$. The error of the exponent is computed as the standard deviation of the MLE \citep{Deluca2013}. Numbers in parentheses correspond to the maximum and minimum value of the amplitude in  dB for each sample. $\hat{n}$ is the number of data entering into the fit and $N$ is the total number of events in the dataset. }
\label{tab:1}
\end{table}

\subsection{Global Fit}
\label{seq:globalfita}
Once the particular fits have been performed, the ranges for which the power-law hypothesis cannot be rejected are known for each experiment $\left[ A_{\rm min_{i}},A_{\rm max_{i}}\right]$ (see Table \ref{tab:1}). For each catalog $i$, we have $\hat{n}_{i}$ events that follow the distribution in Eq. (\ref{eq:ampdis}) and we assume that there exists a global exponent $\hat{\alpha}_{g}$ that characterizes a global distribution that includes the power-law regimes for all the experiments. 

Under these assumptions, for the particular case of amplitudes in  dB, the general log-likelihood function in Eq. (\ref{eq:general}) reads:
\begin{equation}
\begin{split}
\log \mathcal{L} =& \sum_{i=1}^{n_{\rm cat}} \sum_{l=A_{\rm min_{i}}}^{\rm A_{max_{i}}}  \omega_{il} \log  f^{(i)}_{\mathcal{Y}}\left( g(l) \right) = \\
&  \sum_{i=1}^{n_{\rm cat}} \sum_{l=A_{\rm min_{i}}}^{\rm A_{max_{i}}}  \omega_{il} \log  P\left( g\left(l-\Delta\right) \leq \mathcal{V}_{i} < g\left(l+\Delta \right) \right)  
\end{split}
\label{eq:mleglobalAmplitude}
\end{equation}
where $\omega_{il}$ is the number of events with amplitude in  dB $l$ in the $i$-th experiment from a set of $n_{cat}$ catalogs $\left( \sum_{l=A_{\rm min_{i}}}^{A_{\rm max_{i}}}  \omega_{il} = \hat{n}_{i}   \right)$. 

%Note that maximum likelihood estimation would give
%the same results if we were using the 
After maximization of the log-likelihood, 
next step consists in determining whether the null hypothesis of considering a global exponent $\hat{\alpha}_{g}$ is compatible with the values of the particular fits shown in Table \ref{tab:1}. This procedure is explained in more detail in Appendix \ref{sec:AP4}.
\begin{figure}
\includegraphics[scale=0.75]{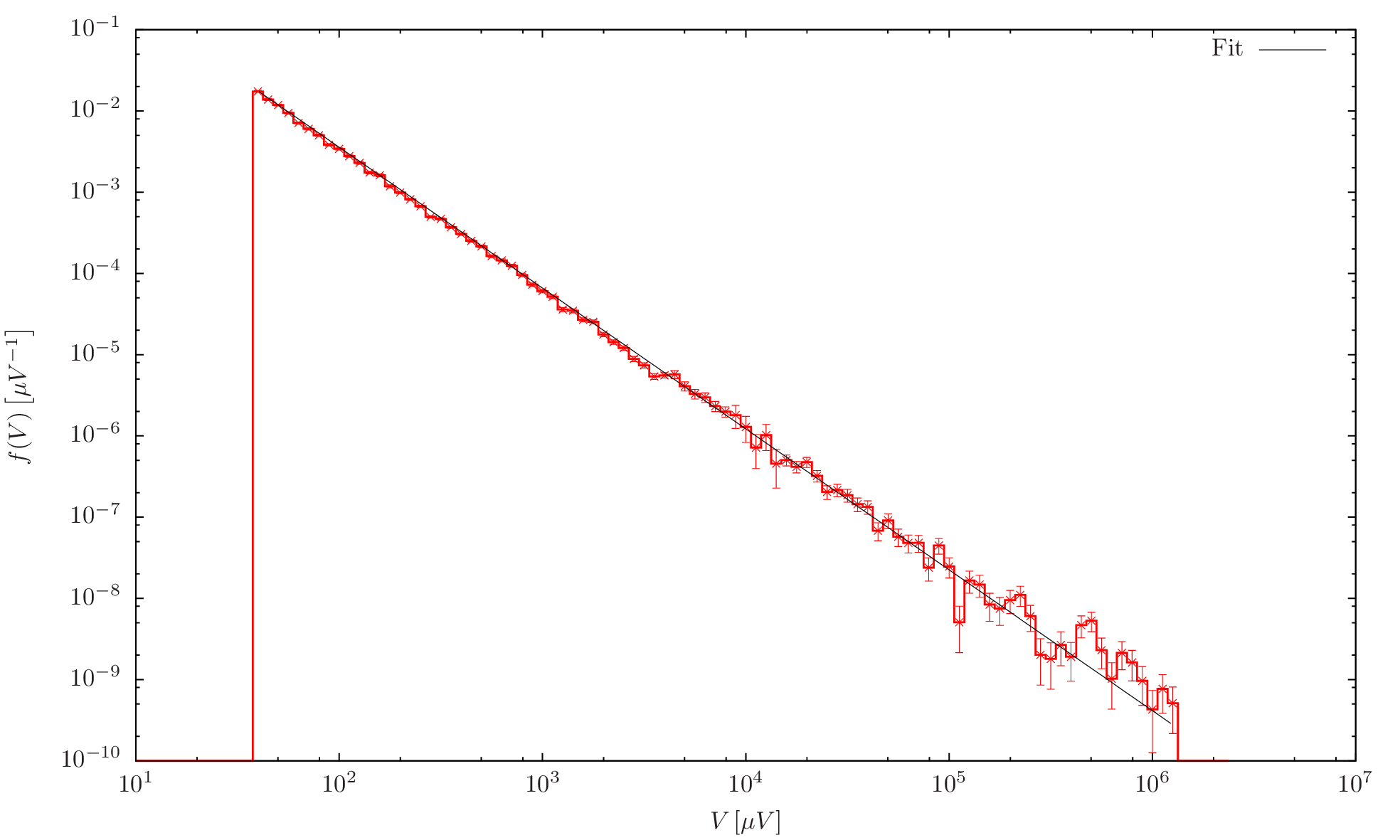}
\caption{\label{fig:globalhista} Aggregated amplitude probability density 
%in $\mu V$ 
of the global distribution with exponent $\alpha_{g}=1.740\pm 0.006$, $p$-value$=0.36$, and number of fitted data $\mathcal{N}=31441$. Error bars are estimated as the standard deviation for each bin \citep{Deluca2013}. 
Black solid line shows the fit of a truncated power-law with exponent $\hat{\alpha}_{g}=1.740$
and ranging almost $5$ decades, from $g(32 -\Delta)$ to $g(122 +\Delta)$. }
\end{figure}
The global fit yields a global exponent $\hat{\alpha}_{g}=1.740 \pm 0.006$ with a $p$-value$=0.36$ for $\mathcal{N}=\sum_{i=1}^{n_{\rm cat}}\hat{n}_{i}=31441$ events. Note that the value of the global exponent is in agreement with the weighted harmonic mean 
$$
\hat{\alpha}_{g}= 1.740 \simeq 1+\frac{\mathcal{N}}{\sum_{i=1}^{n_{\rm{cat}}} \frac{\hat{n}_{i}}{\hat{\alpha}_{i}-1}}= 1.741,
$$ 
see Appendix \ref{sec:AP5} for a justification of this result. This procedure has been tested over simulated power-law data with the same parameters as in Table \ref{tab:1}, yielding  acceptable $p$-values.

Figure \ref{fig:globalhista} shows the global PDF for the amplitudes and the global fit.  
Observe how the global exponent is valid along $4.5$ orders of magnitude, giving an unprecedented broad fitting-range in amplitudes.
The procedure to construct this aggregated histogram is explained in Appendix \ref{sec:AP2}.
As the estimation of the probability density is done using bins \citep{Deluca2013}, 
note that
one can safely replace the unknown values of the random variable $\mathcal{V}$ 
by the known discretized values of $\mathcal{Y}$.
The only requirement is that the width of the bins is not smaller than the discretization of 
$\mathcal{Y}$.
\section{Energies}
\label{seq:energy}
\subsection{Particular fits}
\begin{figure}
\includegraphics[scale=0.65]{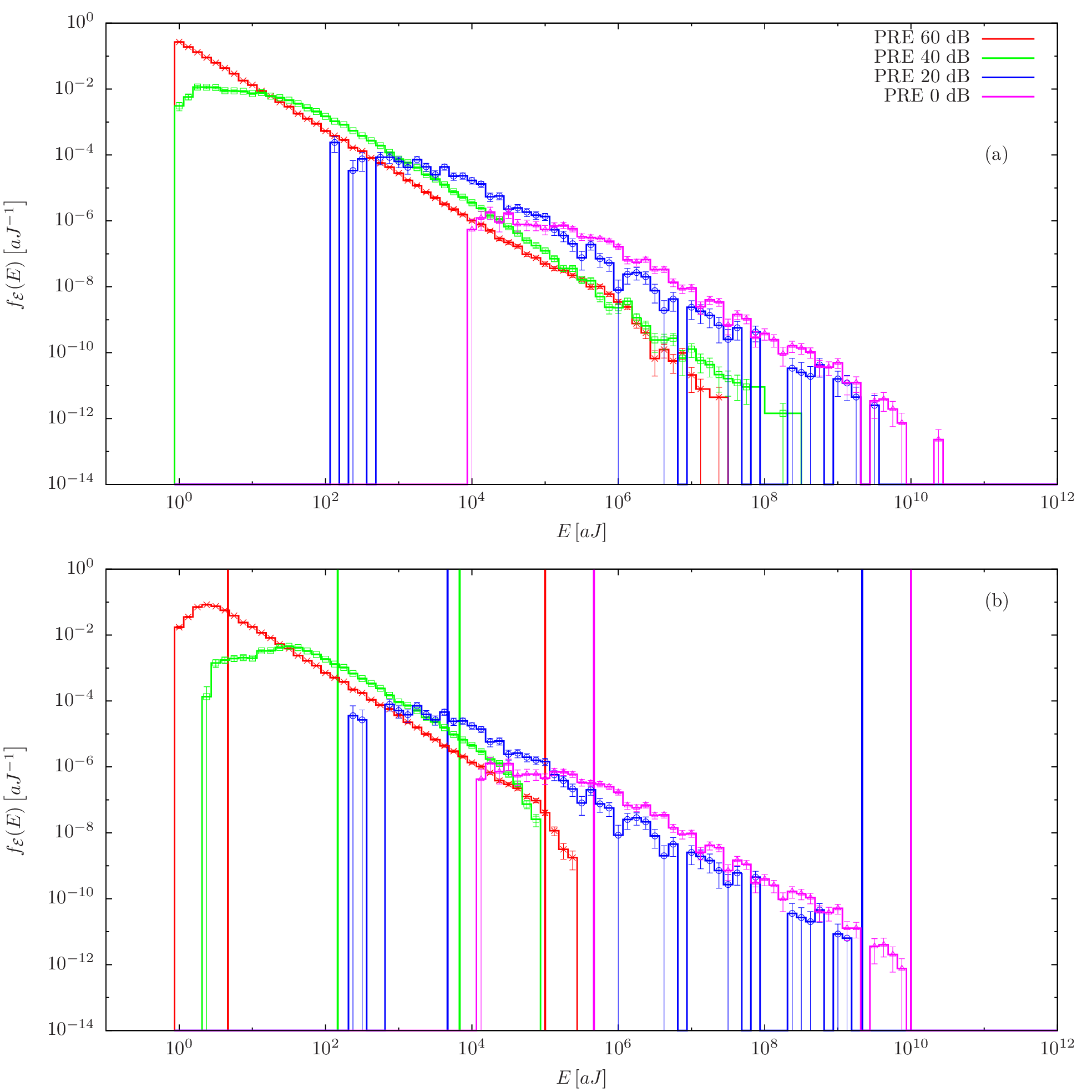}
\caption{\label{fig:rawe} Estimated energy PDFs for the different experiments performed at different pre-amplifications (PRE). (a) Complete datasets. (b) Events with amplitude in the power-law range  $V \in \left[ V_{\rm min_{i}}, V_{\rm max_{i}} \right]$. Vertical bars of the same color as the PDFs correspond to the power-law ranges exposed in Table \ref{tab:tab2}. Error bars are estimated as the standard deviation for each bin \citep{Deluca2013}.}
\end{figure}

Figure \ref{fig:rawe} (a) shows the energy distributions for the complete dataset of all the experiments performed at different values of the pre-amplification. 
%\subsection{Particular fits}
Contrarily to the case of amplitudes, continuous values of the energy are collected 
(see Fig. \ref{fig:rawe}). Due to the problems of saturation for large amplitudes and the presence of noise for small amplitudes, the energy corresponding to these events is not well estimated. In the following analysis we only consider events whose amplitude lies in  $\left[ V_{\rm min_{i}}, V_{\rm max_{i}} \right]$, where the ranges are those that have been found in the particulars fits for amplitude PDF in Table \ref{tab:1} (see Fig. \ref{fig:rawe} (b)). We propose that the energy follows a truncated continuous power-law PDF:
\begin{equation}
f_{\mathcal{E}}(E) dE=\frac{1-\epsilon}{E_{\rm max}^{1-\epsilon}-E_{\rm min}^{1-\epsilon}}E^{-\epsilon} dE.
\label{eq:energy}
\end{equation}
By fixing the values of the range $\left[ E_{\rm min}, E_{\rm max}  \right]$ we find the value of $\epsilon$ that maximizes the next log-likelihood function:
\begin{equation}
\log \mathcal{L} = n \log \left( \frac{1-\epsilon}{E_{\rm max}^{1-\epsilon}-E_{\rm min}^{1-\epsilon}}   \right) - \epsilon \sum_{j=1}^{n} \log E_{j},
\end{equation}
where $E_{j}$ are the particular values of the energy and $n$ is the number of data in $\left[ E_{\rm min}, E_{\rm max} \right]$. Note that $E_{\rm min}$ and $E_{\rm max}$ do not have a direct correspondence with $V_{\rm min}$ and $V_{\rm max}$. Explicit details of this particular fit are exposed in Appendix \ref{sec:AP1}. The values of the fitted exponent $\hat{\epsilon}$ for different values of $E_{\rm min}$ and $E_{\rm max}$ are shown in Appendix \ref{sec:AP3} by using MLE exponent maps \citep{Baro2012}. In Table \ref{tab:tab2} we present fitting parameters when the minimum significance level is set as $p_{c}=0.20$. 
The values of the exponents are in rough agreement with the one reported in Ref. \citep{Baro2013}, 
in particular the value for 60 dB.
\begin{table}[htbp]
%\begin{tabular}{|r|r|r|r|r|r|}
\begin{tabular}{| l | r | rr | rr | rr | r |c| }
\hline
\multicolumn{1}{| l |}{PRE in dB} & \multicolumn{1}{| c |}{$\hat{\epsilon}$} & \multicolumn{2}{| c |}{$E_{\rm min}$ in aJ}  & \multicolumn{2}{| c |}{$E_{\rm max}$ in aJ} & \multicolumn{2}{| c |}{$\hat{n}(N)$} & \multicolumn{1}{| c |}{$p$-value}  \\ \hline
60 & $1.360\pm 0.004$  & $4.642$ & $(1.001)$ & $10^{5}$ & $(3.005 \times 10^{5})$ & 16342 &(21414) & $0.43$  \\ \hline
40 & $1.32 \pm 0.02$ & $146.780$ & $(3.133)$ & $6.812 \times 10^{3}$ & $(9.378 \times 10^{4})$ & 4814 &(9146) & $0.32$  \\ \hline
20 & $1.29\pm 0.02$ & $4641.589$ & $(270.446)$ & $2.15 \times 10^{9}$ & $(1.588 \times 10^{9})$ & 284 &(353) & $0.33$ \\ \hline
0 & $1.27 \pm 0.02$ & $4.642 \times 10^{5}$ & $(1.545 \times 10^{4})$ & $10^{10}$ & $(9.156 \times 10^{9})$ & 396 &(528) & $0.55$ \\ \hline
\textbf{Global} & 1.352 $\pm$ 0.004 & $4.642$ & $(1.001)$ & $ 10^{10}$ & $(9.156 \times 10^{9})$ & 21836 & (31441) & 0 \\ \hline
\end{tabular}
\caption{Fitted parameters for Eq. (\ref{eq:energy}) for each experiment. $\hat{\epsilon}$ corresponds to the fitted exponent in the range $\left[ E_{\rm min},E_{\rm max} \right]$ for which the goodness-of-fit test exceeds the significance level $p_{c}=0.2$. $\hat{n}$ is the number of data entering into the fit and $N$ is the total number of events in the dataset. Numbers in parentheses refer to 
values of the energy in the range
%the complete dataset of energies with $V \in 
$\left[ V_{\rm min_{i}}, V_{\rm max_{i}} \right]$. 
Error bars of the exponent correspond to the standard deviation of the MLE.}
\label{tab:tab2}
\end{table}
\subsection{Global fit}
In order to write the log-likelihood function of the global fit, one has to consider that each experiment contributes with $\hat{n}_{i}$ data which are distributed according to Eq. (\ref{eq:energy}) in the range $\left[ E_{\rm min_{i}},E_{\rm max_{i}} \right]$ with a global exponent $\epsilon_{g}$:
\begin{equation}
\log \mathcal{L} = \mathcal{N} \log \left(  1-\epsilon _{g} \right) - \epsilon_{g} \sum_{i=1}^{n_{\rm cat}} \sum_{j=1}^{\hat{n}_{i}} \log E_{ij} - \sum_{i=1}^{n_{\rm cat}} \hat{n}_{i} \log \left( E_{\rm max_{i}}^{1-\epsilon_{g}} - E_{\rm min_{i}}^{1-\epsilon_{g}}   \right)
\end{equation}
where  $\hat{n}_{i}$ is the number of data in the $i$-th catalog, $\mathcal{N}=\sum_{i}^{n_{\rm cat}} \hat{n}_{i}$ and $E_{ij}$ are the values of the energy in each power-law regime $i$. The values of the ranges $\left[ E_{\rm min_{i}},E_{\rm max_{i}} \right]$ are taken from the particular fits in Table \ref{tab:tab2}. Details of the goodness-of-fit test for this global fit are explained in Appendix \ref{sec:AP4}.
By considering the particular ranges shown in Table \ref{tab:tab2}, the global fit of the energy exhibits an exponent $\hat{\epsilon}_{g}=1.352\pm 0.004$ ($\mathcal{N}=\sum_{i=1}^{n_{\rm cat}} \hat{n}_{i}=21836$) along more than nine decades. As it happens for the case of the global amplitude distribution, the value of the global exponent is in agreement with the weighted harmonic mean $$\hat{\epsilon}_{g}= 1.352 \simeq 1+\frac{\mathcal{N}}{\sum_{i=1}^{n_{\rm{cat}}} \frac{\hat{n}_{i}}{\hat{\epsilon}_{i}-1}}=1.347.$$ 
As we have mentioned, this result is justified in Appendix \ref{sec:AP5}. 
%for a justification of this result.  

\begin{figure}
\includegraphics[scale=0.75]{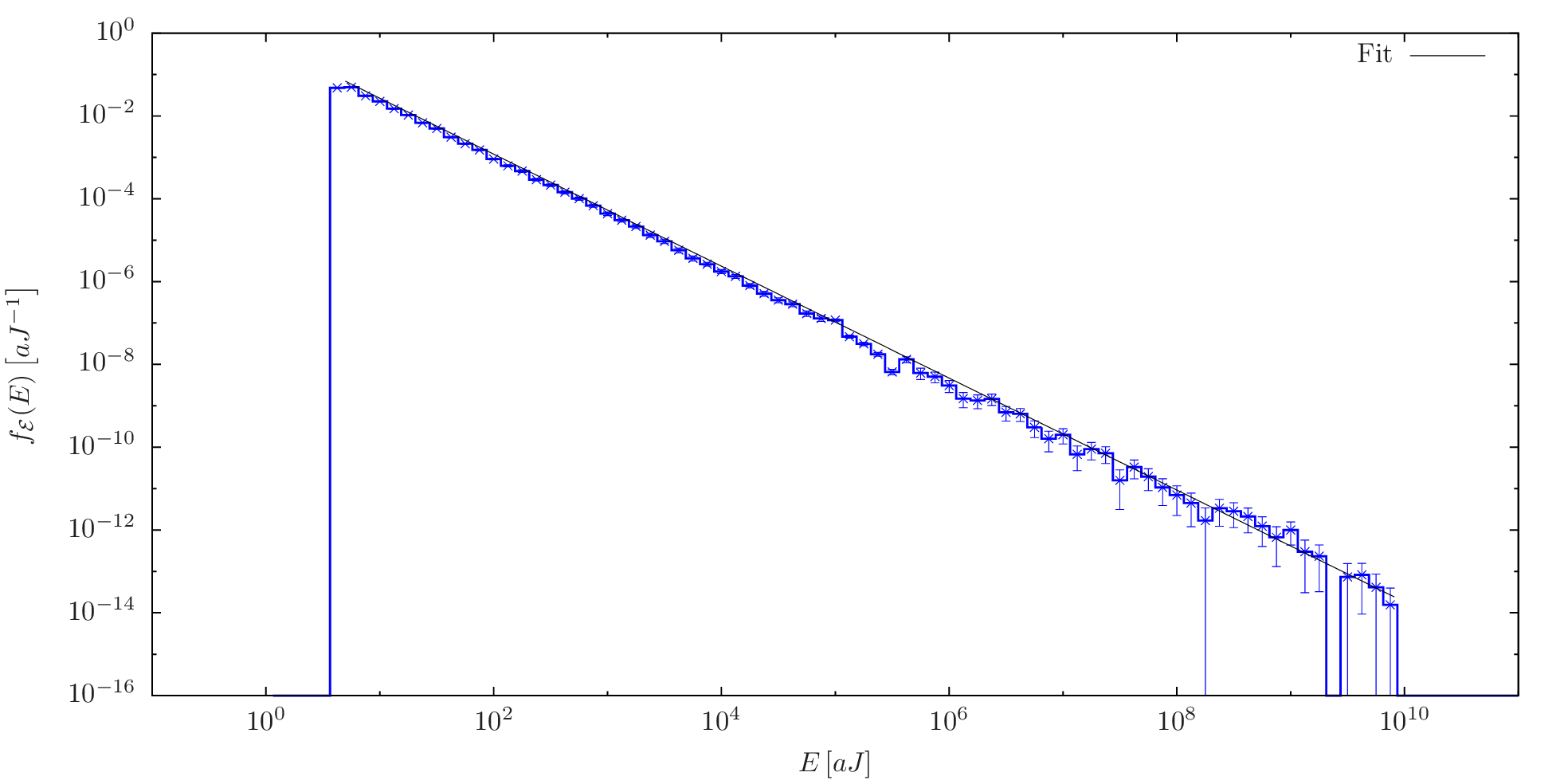}
\caption{\label{fig:globale} Aggregated empirical energy PDF for all the experiments.  
Error bars are estimated as the standard deviation for each bin \citep{Deluca2013}. Black solid line corresponds to the fit of the truncated power-law with exponent $\hat{\epsilon}_{g}=1.352$ in the range  $\left[ 4.642 \mbox{ aJ} , 10^{10} \mbox{ aJ} \right]$. }
\end{figure}

Nevertheless, this global fit does not fulfil the goodness-of-fit test and the null hypothesis $\rm H_{0}$ that all the catalogs share a common exponent $\hat{\epsilon}_{g}$ is rejected.
In Fig. \ref{fig:globale} we show the aggregated empirical probability density for the energy of the AE events. This histogram has been constructed following the procedure explained in Appendix \ref{sec:AP2}. Simulated data with the same parameters as in Table \ref{tab:tab2} also yield the same rejection of the null hypothesis. 
We have performed the same analysis without the restriction of just considering events whose amplitude $V \in \left[ V_{\rm min_{i}}, V_{\rm max_{i}} \right]$ as well as sparing some catalogs. In all the cases, the rejection of the null-hypothesis occurs.

This result could be explained by a biased measurement of the energy caused by the interplay between the measured event duration and the detection threshold. The higher the threshold, the shorter the duration and the lower the energy. This fact would not be significant for the case of the amplitudes, since these are independent of the duration, but it should be for the energy since it corresponds to the integrated (squared) AE signal along the registered duration.

\section{Conclusions}
\label{Conclusions}
In this work we have presented a methodology to estimate a global exponent for the PDF of certain avalanche observables by using different catalogs of events. This methodology has been applied to amplitudes and energies in AE avalanches recorded during different compression experiments of porous glasses. For the case of the amplitude PDF, a global exponent has been found spanning $4.5$ orders of magnitude. To our knowledge, this is the broadest fitting range that has been found for the amplitude distribution of AE events. For the case of the energies, we graphically obtain an apparent power-law spanning $9.5$ decades. However, precise statistical analysis shows that the hypothesis of the existence of a global exponent does not hold. Experimental limitations due to the set of thresholds and definitions of AE avalanches could justify the rejection of the null hypothesis. We expect this methodology to be useful to broaden the range of power-law fits in different distributions that appear in experimental works in condensed matter physics and in other complex systems, for instance, earthquakes.
\begin{acknowledgments}
The research leading to these results has received founding from ``La Caixa" Foundation.
We also acknowledge financial support 
through the ``Mar\'{\i}a de Maeztu'' Programme for Units of Excellence in R\&D (MDM-2014-0445), as well as from projects
FIS2015-71851-P, MAT2013-40590-P, and the
Proyecto Redes de Excelencia 2015 MAT 2015-69-777-REDT, 
all of them from the Spanish Ministry of Economy and Competitiveness,
and from 2014SGR-1307 from AGAUR.   
\end{acknowledgments}
%\newpage
\appendix
\section{Fitting a truncated power-law distribution to a dataset}
\label{sec:AP1}
In this appendix we present the procedure to fit via maximum likelihood estimation (MLE) a truncated power-law probability density function (PDF) to an empirical dataset of $N$ values $\lbrace x \rbrace$. The procedure is essentially the same as the one in Ref.\citep{Deluca2013}, but we summarize it here for completeness.  The power-law PDF $f\left( x; \gamma, a,b\right)$ is characterized by an exponent $\gamma$ and two values for the lower $a$ and upper $b$ truncations. 
The first step consists in writing the log-likelihood function:
\begin{equation}
\log \mathcal{L} = \log  \prod_{i=1}^{n} f\left(x_{i};\gamma , a,b \right) = \sum_{i=1}^{n} \log f\left(x_{i};\gamma , a,b \right),
\label{eq:likelihood}
\end{equation}
where $n$ is the number of data between $a$ and $b$. Depending on whether data is continuous or discrete, one has to use a different expression of $f(x;\gamma,a,b)$. Since the values of the fitting range $a$ and $b$ are not known \textit{a priori}, twenty partitions per decade in log-scale are used in order to sweep all the possible intervals $\left[ a,b \right]$ for the amplitude and six partitions per decade for the energy. The value of the empirical exponent $\gamma$ that maximizes Eq. (\ref{eq:likelihood}) is computed for each interval by means of the function \textit{``optimize"}, which is already implemented in R programming language \citep{RDocumentation,Brent2002}. Once this value is found, one has to determine through a statistical test whether the fit is ``acceptable" or not. The null hypothesis states that data $\lbrace x \rbrace$ is sampled from a truncated power-law distribution with exponent $\gamma$. In order to check whether this null hypothesis is rejected or not, the Kolmogorov-Smirnov (KS) distance is used \cite{Press}. 
The KS statistic measures the maximum distance between the empirical cumulative distribution function (CDF) $F_{e}$ (the subscript $e$ refers to empirical) and the analytic expression $F$:
\begin{equation}
d_{e}= \max \big\vert F_{e}\left(  x; a,b \right) -  F\left( x; \gamma, a, b \right)  \big\vert.
\label{eq:ks}
\end{equation}    
Once the value of this distance is known, it is necessary to state whether its value is large or not compared to those $\lbrace d_{\rm sim} \rbrace$ found when the original data is really sampled from a truncated power-law distribution with exponent $\gamma_{e}$. In order to perform this estimation, a number  $N_{\rm sim}  (N_{\rm sim}=1000)$ of simulations of $\hat{n}$ values sampled from $f\left( x; \gamma,a,b \right)$ is done. A fitted exponent $\gamma_{\rm sim}$ for simulated data is computed by maximizing Eq. (\ref{eq:likelihood}) and a KS distance $d_{\rm sim}$ is found for each simulated dataset.
The $p$-value of the test is estimated as the fraction of observations where  $d_{e} \leq d_{\rm sim}$. If the $p$-value of the fit exceeds a certain threshold then one considers that the null hypothesis is accepted (in the sense that it cannot be rejected). 
From all the possible intervals whose $p$-value exceeds $p_{c}$, the one with largest number of data is chosen to yield the right power-law range with exponent $\hat{\gamma}$. 
\section{Goodness-of-fit test for global distributions}
\label{sec:AP4}
Determining whether the null hypothesis of considering a global exponent $\Gamma$ is compatible with the values of the particular fits has some differences with respect to the case explained in Appendix \ref{sec:AP1}. In this appendix we expose the goodness-of-fit test that has been used for the global fit.
First of all, we need to redefine the KS distance in this case. When the value of the global exponent $\hat{\Gamma}$ has been found, one can understand that each dataset that contributes to the global PDF has been fitted with a global exponent $\hat{\Gamma}$ in their particular ranges $\left[ a_{i} ,b_{i}  \right]$ ($i=1,...,n_{\rm cat}$). Therefore $n_{\rm cat}$ KS distances can be found by :
\begin{equation}
D_{e,i}= \max \biggr\vert F_{e,i}\left(  x; a_{i},b_{i}  \right) - F_{i}  \left(x;\hat{\Gamma},a_{i},b_{i}  \right)  \biggr\vert,
\end{equation}
where the subindex $i$ refers to the $i$-th catalog, $F_{e}$ is the empirical CDF and $F$ is the analytical CDF.
In order to compute a global KS distance, we perform the following summation:
\begin{equation}
D=\sum_{i=1}^{\rm n_{\rm cat}} \sqrt{ \hat{n}_{i}} D_{e,i},
\end{equation}
where the factors $\sqrt{ \hat{n}_{i}}$ are due to the scaling of the KS distance
with the number of data \cite{Press}.
Once the empirical KS distance is found, one needs to determine whether this distance is big or small in relation to the KS distance found for data sampled from a PDF with the same parameters $a_{i}$,$b_{i}$, $\hat{\Gamma}$ and $\hat{n}_{i}$.
Data is generated in the range given by the particular fit of the $i$-th catalog with probability $q_{i}=\hat{n}_{i}/N$, where $N=\sum_{i=1}^{\rm n_{cat}} \hat{n}_{i}$. Note that the particular number in each simulated dataset is not necessarily the empirical one $\hat{n}_{i}$ but the total number of data $N$ is maintained. Hence, one needs a first random number to choose the dataset $i$ and therefore the range $\left[ a_{i},b_{i} \right]$ and a second one to generate the random truncated power-law number in that range with exponent $\hat{\Gamma}$\citep{Deluca2013}. When $N$ events have been generated according to this procedure, one finds the global exponent $\hat{\Gamma}_{\rm sim}$ by maximizing the global log-likelihood and computes the global KS distance $D_{\rm sim}$ for simulated data. 
By performing several realizations of the previous procedure, one can estimate the $p$-value of the fit by computing the fraction of simulated datasets where the simulated global KS distance is larger than the empirical one.
\section{MLE Exponent Maps}
\label{sec:AP3}
In order to complement the information of the particular fits, we show the MLE exponent maps. These maps show the value of the exponent of a truncated power-law as a function of the values of the upper and lower truncations. This kind of representation is useful since it gives information about how stable is the value of the exponent as the truncations of the power-law fit change.
\begin{figure}
\includegraphics[scale=0.625]{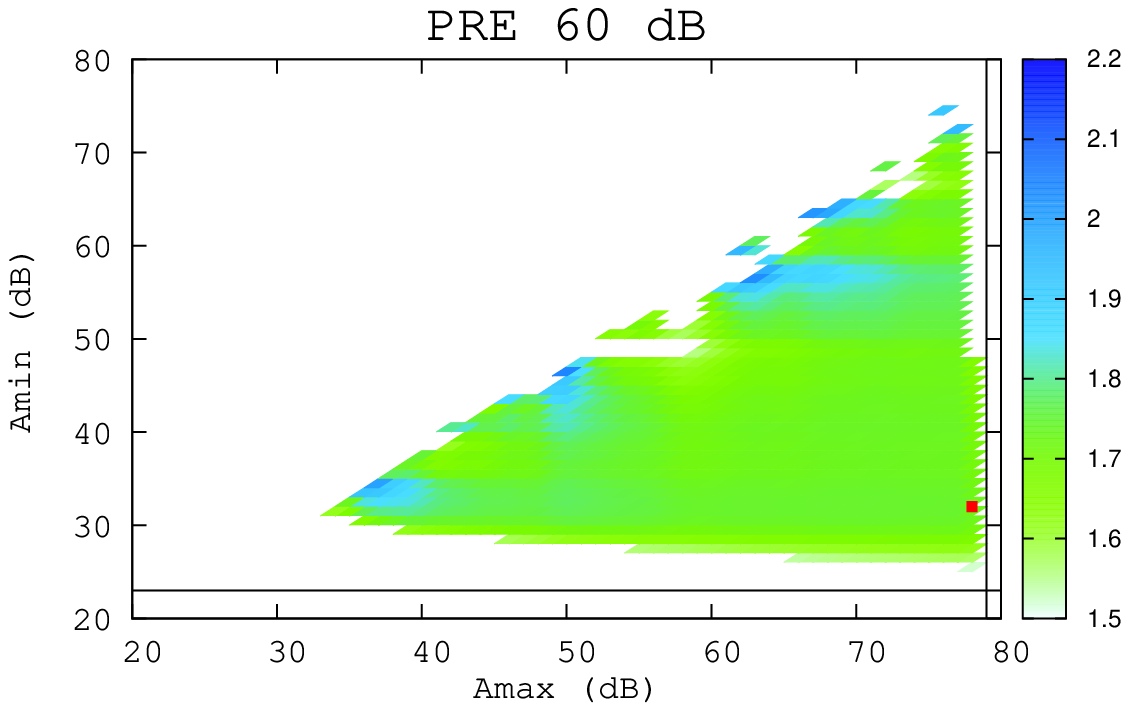}
\includegraphics[scale=0.625]{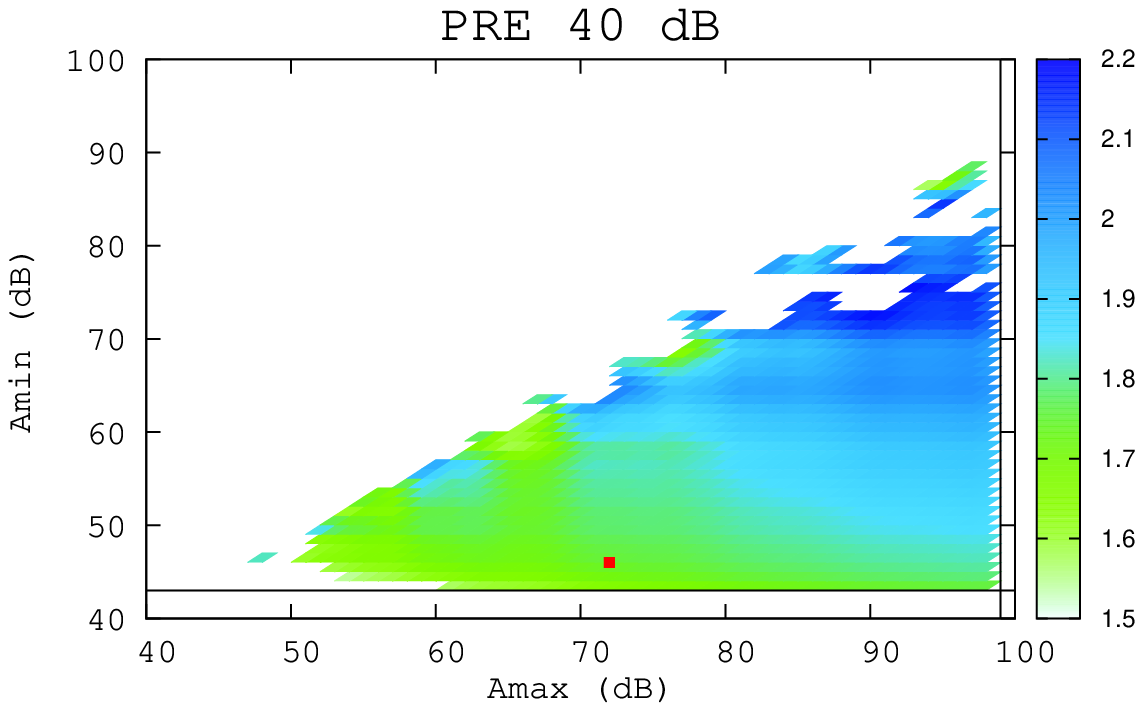}
\includegraphics[scale=0.625]{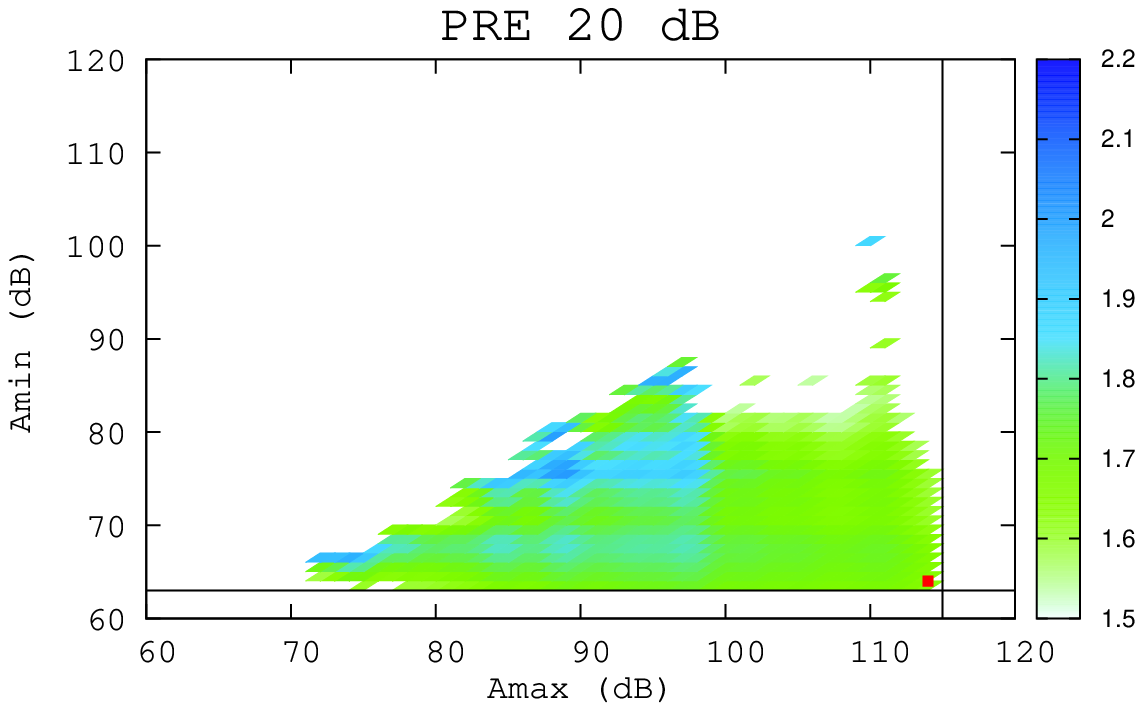}
\includegraphics[scale=0.625]{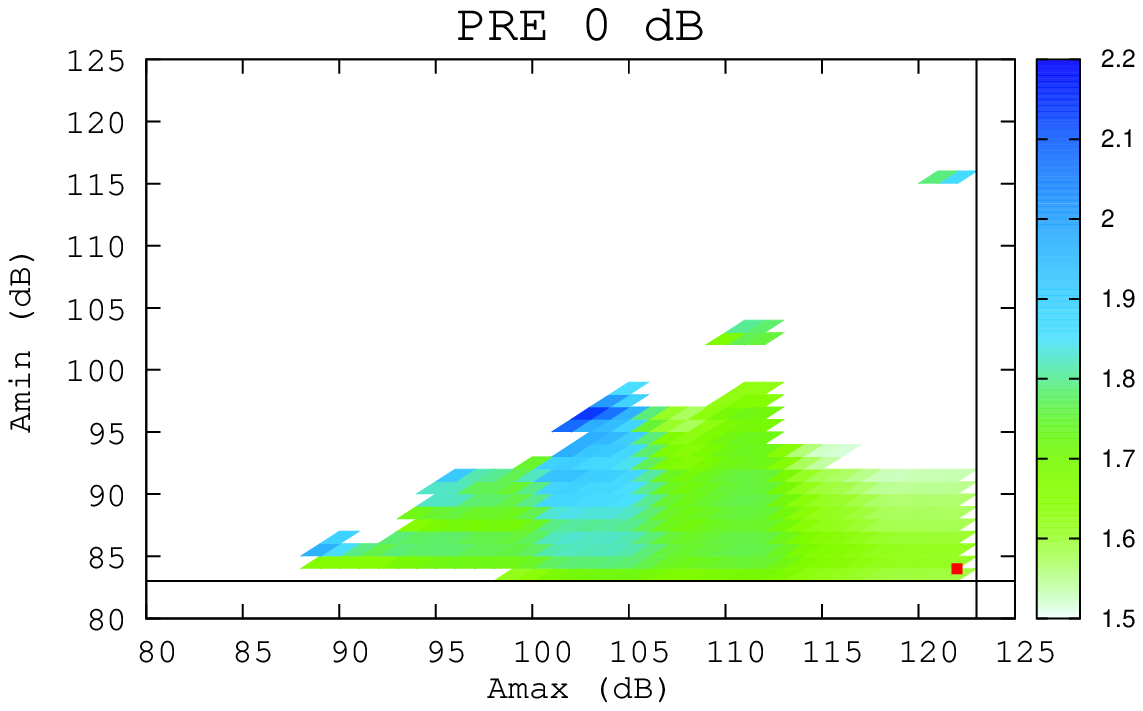}
\caption{\label{fig:MLEAMP} MLE Exponent Maps for the distribution of amplitudes in  dB of every experiment. Red points indicate the values of the exponent and the upper and lower truncations found in Table \ref{tab:1} in the main text.}
\end{figure}
In Figs.\ref{fig:MLEAMP} and \ref{fig:MLEE} we show the MLE exponent maps for the case of the amplitudes in  dB and energy respectively. Due to the condition that the variable $\mathcal{X}$ must fulfil $x_{min} \leq x_{max}$ for the upper and lower truncations, the maps exhibits a triangular shape. White gaps correspond to regions that are out of the color range that appears at the right of each map. Black solid lines correspond to the maximum and minimum values in the sample. The points in each map show the upper and lower truncations as well as the exponent of the fit which has been done according to the fitting procedure explained in Appendix \ref{sec:AP1}. As it can be observed, this fit is placed in uniform-coloured regions of the map. 
\begin{figure}
\includegraphics[scale=0.625]{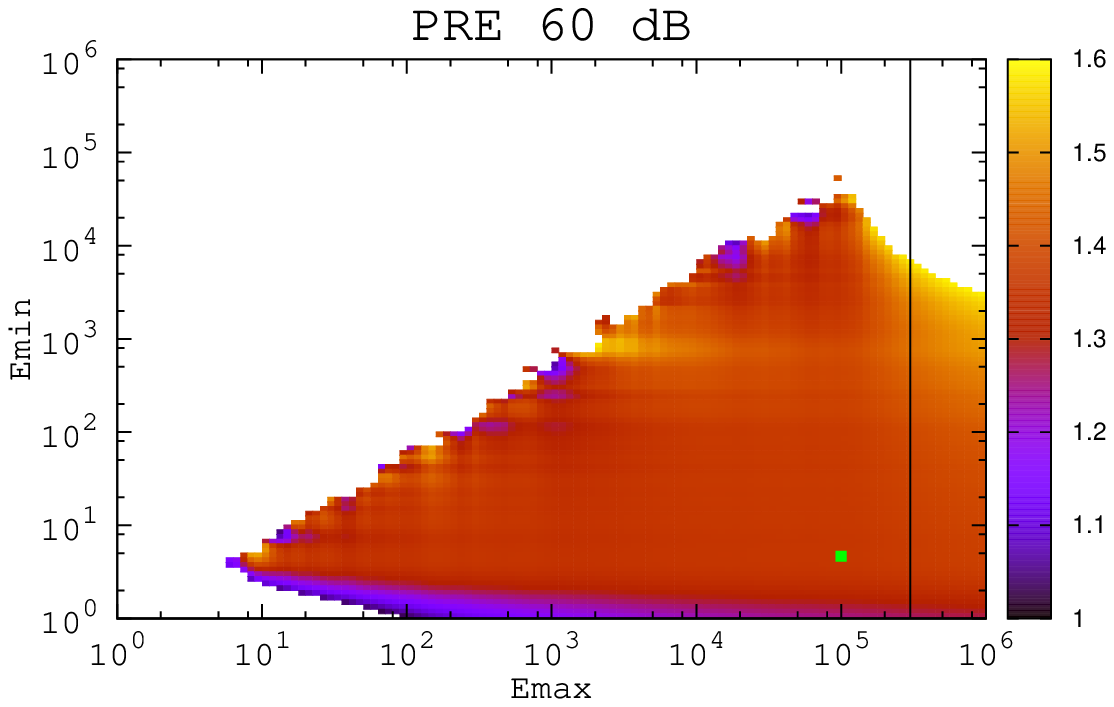}
\includegraphics[scale=0.625]{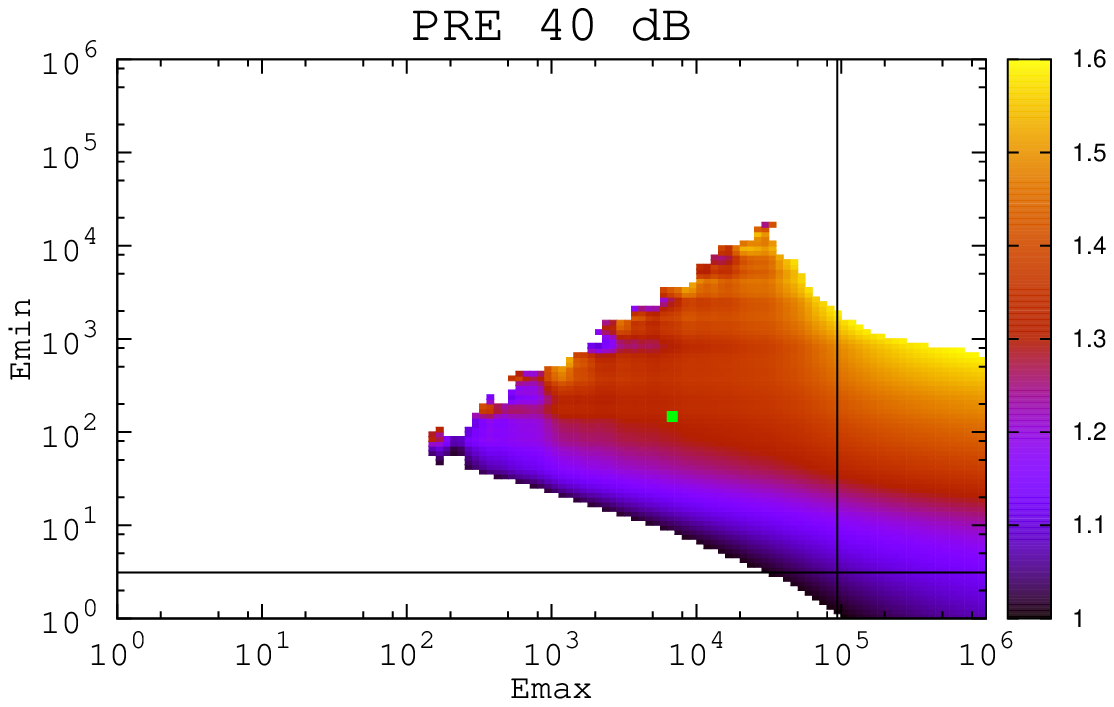}
\includegraphics[scale=0.625]{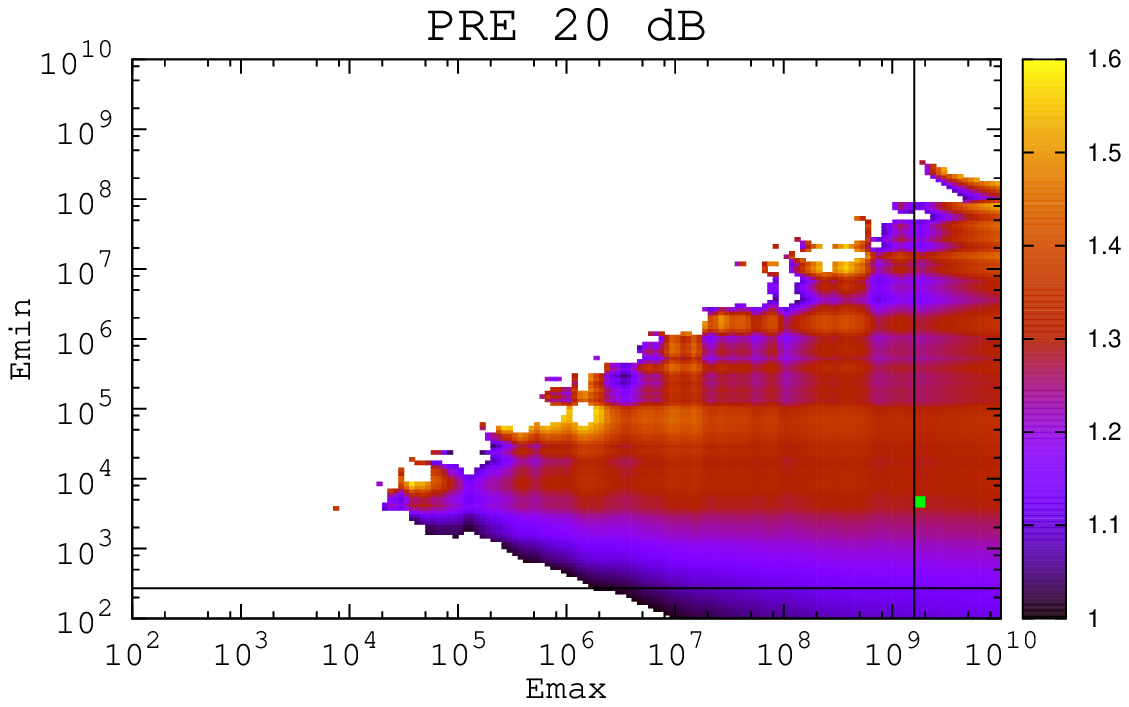}
\includegraphics[scale=0.625]{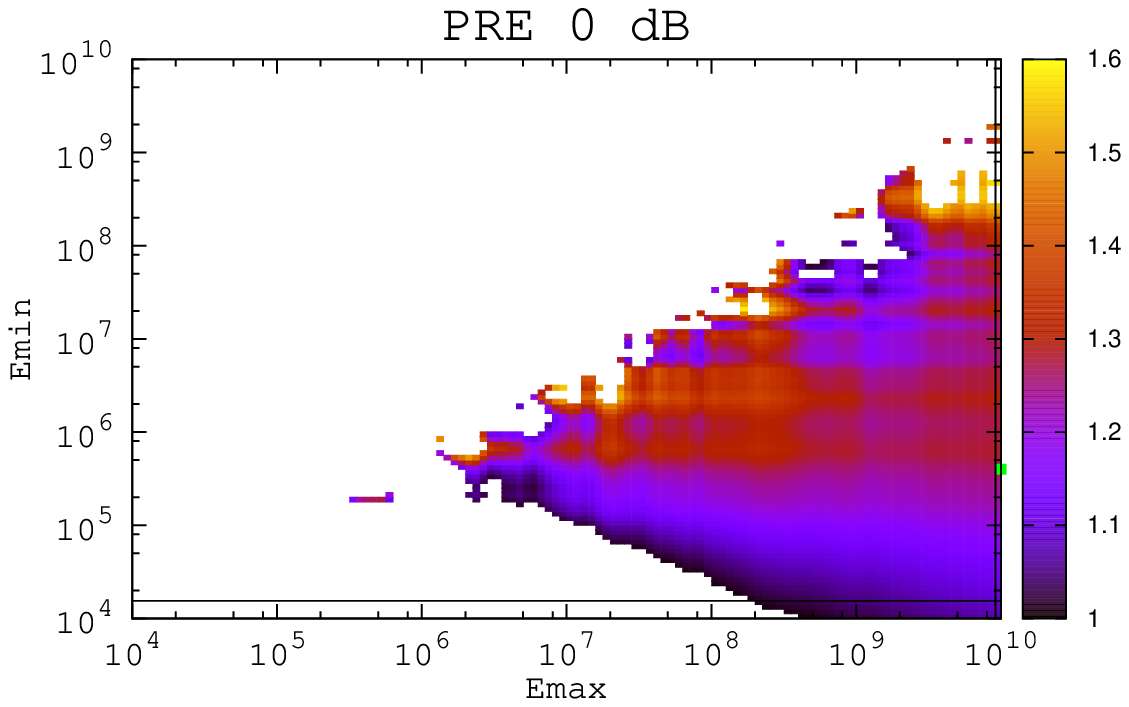}
\caption{\label{fig:MLEE} MLE Exponent Maps for the energy for each experiment. Green points indicate the values of the exponent and the upper and lower truncations found in Table \ref{tab:tab2} in the main text.} 
\end{figure}

\section{Global exponent}
\label{sec:AP5}
In this appendix we develop Eq. (\ref{eq:general}) for a non-truncated power-law PDF in order to relate the global exponent with the particular ones. Let us consider $n_{\rm cat}$ catalogs with $\hat{n_{i}}$ events characterized by a variable $\mathcal{X}$. We propose that for the $i$-th catalog, data follows a power-law with exponent $\gamma$ from a lower cut-off $a_{i}$ to $\infty$:
\begin{equation}
f_{\mathcal{X}}^{(i)}(x) = \frac{\gamma-1}{a_{i}^{1-\gamma}}x^{-\gamma}
\label{eq:ap51}
\end{equation}
By finding the values of the particular exponents that maximize the log-likelihood expression in Eq. (\ref{eq:likelihood}) we obtain \citep{Deluca2013,Clauset2009}:
\begin{equation}
\hat{\gamma}_{i} = 1 + \frac{\hat{n}_{i}}{\sum_{j=1}^{\rm \hat{n}_{i}} \log\left( x_{j}/a \right)}
\label{eq:ap52}
\end{equation}
Now we consider that data in these $n_{\rm cat}$ catalogs follow PDFs with different cut-offs $a_{i}$ but sharing a global exponent $\Gamma$. If we bring these PDFs to the global log-likelihood fo Eq. (\ref{eq:general}), we obtain:
\begin{equation}
\log \mathcal{L} = \sum_{i=1}^{n_{\rm cat}} \hat{n}_{i}\log\left( \Gamma-1 \right) + \sum_{i=1}^{n_{\rm cat}}\hat{n}_{i}\left( \Gamma-1 \right) \log a_{i} - \Gamma \sum_{i=1}^{n_{\rm cat}}\sum_{j=1}^{\hat{n}_{i}} \log x_{j} 
\end{equation}
By deriving this expression with respect $\Gamma$, we can obtain the value of the global exponent that maximizes the log-likelihood:
\begin{equation}
\hat{\Gamma} = 1 + \frac{\mathcal{N}}{\sum_{i=1}^{\rm n_{cat}} \sum_{j=1}^{\rm \hat{n}_{i}} \log\left( x_{j}/a_{i} \right) }
\end{equation}
We can relate this global exponent $\Gamma$ with the particular exponents from Eq. (\ref{eq:ap52}) leading to:
\begin{equation}
\frac{\mathcal{N}}{\hat{\Gamma}-1} = \sum_{i=1}^{\rm n_{cat}} \frac{\hat{n}_{i}}{\hat{\gamma}_{i}-1} 
\end{equation}
Hence, for the case of non-truncated power-law PDFs, the global exponent is related to the weighted harmonic mean of $\gamma-1$, being $\gamma$ the exponent of the particular PDF of the $i$-th catalog.
If the range is sufficiently big, one is able to find an agreement between the values of $\Gamma$ where the particular PDFs are truncated power-laws, as for large ranges one expects the non-truncated power-law solution provides a good approximation.

\section{Aggregated Global Histogram}
\label{sec:AP2}
\begin{figure}
\includegraphics[scale=0.6]{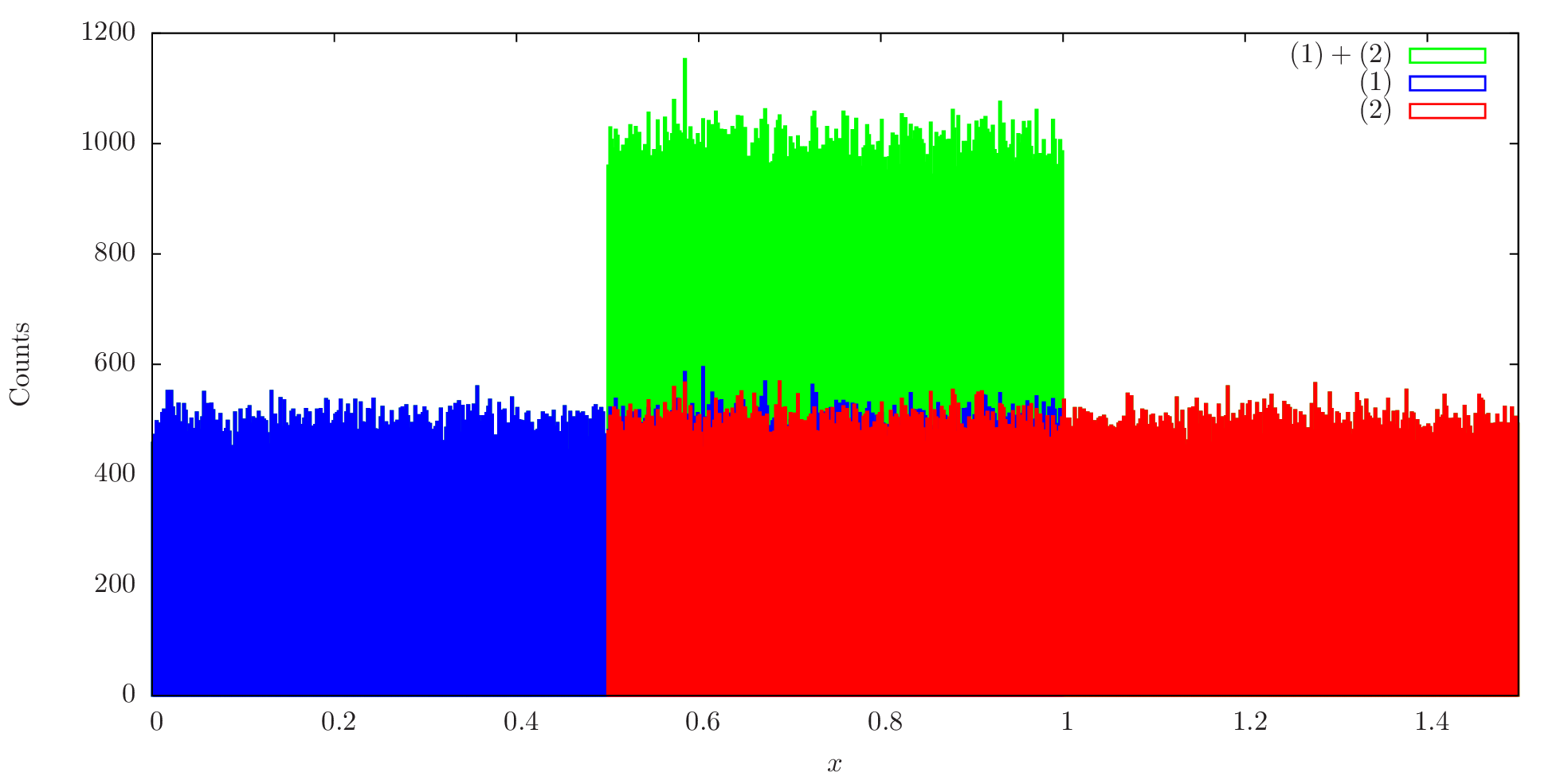}
\caption{\label{fig:histuniform} Histogram of $N=2.5\times 10^{4}$ numbers sampled from a $U(0,1)$ in blue $(1)$, $N=2.5\times 10^{4}$ numbers sampled from a $U(0.5,1.5)$ in red $(2)$. The sum of these histograms  $(1+2)$ is shown in green, where an excess of counts can be appreciated in the intersection region.} 
\end{figure}

We would like to represent a global histogram of a certain variable $x$ which is sampled from $n_{cat}$ datasets with $n_{i}$ number of data for the $i$-th one. Although the underlying distribution can be considered the same for all the datasets, the difference in the number of data and the different domains can lead to bumps and irregularities. Two overlapping histograms are shown in Fig. \ref{fig:histuniform} (a) sampled from a uniform distribution between $0$ and $1$ ($U(0,1)$) for histogram $(1)$ and from a $U(0.5,1.5)$ for histogram $(2)$. We would like to construct a single histogram without the anomalies that appear in the overlapping region.

The methodology is based on assembling recursively datasets by pairs. This procedure is independent on the shape of the histogram. In Fig. \ref{fig:histuniform} we present it with uniform PDFs for simplicity but it could be used for any other distribution. Instead of considering one element as a count we consider that each individual contributes as the inverse of the sample intensity. As in Fig. \ref{fig:histuniform}, let us consider two datasets $(1)$ and $(2)$ and let $n_{1}$ be the number of elements from $(1)$ overlapping with those in $(2)$ and $n_{2}$ the number of elements from $(2)$ overlapping with those in $(1)$. 

The first step consists in considering that non-overlapping elements from $(1)$ and $(2)$ contribute $\frac{1}{n_{1}}$ and $\frac{1}{n_{2}}$ respectively whereas the overlapping elements contribute $\frac{1}{n_{1}+ n_{2}}$ to the global histogram. After this first rescaling, datasets $(1)$ and $(2)$ are assembled in a single histogram $(1+2)$. 

In order to add a new dataset $(3)$, we need to evaluate the contribution of $((1)+(2))$ that overlaps with $(3)$. We define $n_{12}$ as the number of elements of $(1)$ and $(2)$ overlapping with $(3)$ and the weight $p_{12}= \underbrace{ \frac{1}{n_{1}} + \frac{1}{n_{1}} + ... + \frac{1}{\left( n_{1}+n_{2}  \right)} +\frac{1}{\left( n_{1}+n_{2}  \right)} + ... + \frac{1}{n_{2}}+...}_{n_{12} \rm terms}$. 

The second step consists in rescaling the factors that were used to assemble the histogram in the first step by $\frac{n_{12}}{p_{12}}$. In this way, the contribution of the overlapping elements of $((1)+(2))$ with $(3)$ is computed correctly for the next step.

The third step is the same as the first one but considering that non-overlapping elements from the assembled histogram contribute $\frac{1}{n_{12}}$, the non-overlapping elements from the new dataset contribute $\frac{1}{n_{3}}$ and the overlapping elements $\frac{1}{\left( n_{12} + n_{3} \right)}$.

One can iterate this procedure in order to add as many datasets as necessary. In order to have an estimation of a PDF, firstly, rescaled counts should be divided by the bin-width for each bin and, secondly, the resulting histogram has to be divided by its area in order to fulfil normalization.

\bibliographystyle{unsrt}
\bibliography{AE}

\end{document}